\documentclass[12pt,preprint]{aastex}
\usepackage{amsmath}
\usepackage{epsfig}
\usepackage{natbib} 
\bibpunct[,]{(}{)}{;}{a}{}{,} 

\begin{document}
\begin{abstract}

Recent developments in numerical relativity have made it possible
to follow reliably the coalescence of two black holes from near
the innermost stable circular orbit to final ringdown.  This
opens up a wide variety of exciting astrophysical applications
of these simulations.  Chief among these is the net kick received
when two unequal mass or spinning black holes merge.  The magnitude of this
kick has bearing on the production and growth of supermassive
black holes during the epoch of structure formation, and on
the retention of black holes in stellar clusters.  Here we report
the first accurate numerical calculation of this kick, for two
nonspinning black holes in a 1.5:1 mass ratio, which is expected
based on analytic considerations to give a significant fraction of the 
maximum possible recoil.  We have performed multiple runs with different
initial separations, orbital angular momenta, resolutions, extraction
radii, and gauges.  The full range of our kick speeds is
86--116~km~s$^{-1}$, and the most reliable runs give kicks between
86 and 97~km~s$^{-1}$.  This is intermediate between the estimates
from two recent post-Newtonian analyses and suggests
that at redshifts $z\gtrsim 10$, halos with masses 
$\lesssim 10^9\,M_\odot$ will have difficulty retaining coalesced
black holes after major mergers.

\end{abstract}


\title{Getting a kick out of numerical relativity.}
\author{John G. Baker, Joan Centrella, Dae-Il Choi\altaffilmark{1}, 
Michael Koppitz,
James~R.~van~Meter, M.~Coleman Miller\altaffilmark{2}}
\affil{Laboratory for Gravitational Astrophysics, 
NASA Goddard Space Flight
Center, Greenbelt, Maryland 20771}
\altaffiltext{1}{Universities Space Research Association,
10211 Wincopin Circle, Suite 500, Columbia, MD 21044}
\altaffiltext{2}{University of Maryland, Department of Astronomy,
College Park, MD 20742-2421}

\maketitle

\section{Introduction}

When two black holes merge, the gravitational waves they produce
will carry away net linear momentum, barring conditions of special
symmetry (e.g., two equal-mass nonspinning black holes receive no
kick).  The magnitude of the resulting recoil is important in a variety of
astrophysical contexts, including the cosmological evolution of supermassive
black holes \citep{Mer04,BMQ04,Hai04,MQ04,YM04,Vol05,Lib05,MAS05} and the 
growth and retention of intermediate-mass
black holes in dense stellar clusters 
\citep{Tan00,MH02a,MH02b,MT02a,MT02b,MC04,GMH04,GMH06,OL06}.
There is therefore a
long history of analytical estimates of this recoil \citep{Per62,
Bek73,Fit83,FD84,RR89,W92,FHH04,BQW05,DG06}.
However, it has been shown that almost all
of the recoil occurs in the strong gravity regime, inside the
innermost stable circular orbit (ISCO).  This is precisely where
analytical treatments are least reliable.  An accurate estimate of
the recoil kick therefore requires full numerical simulations of the
final phase of the inspiral, merger, and ringdown of the coalescence
of two black holes.

Until recently, numerical simulations were not stable and accurate
enough for such estimates.  
This situation has changed dramatically
in the past year, with several groups developing codes that have
allowed the evolution of binary black hole spacetimes from close
to the ISCO through merger and ringdown.  
Most of these codes utilize techniques 
which allow the black holes, with their inherent singularities, to move 
successfully through the computational domain.  One approach 
\citep{Pretorius2005,Pretorius2006} is to 
excise the singular region in the (physically inaccessible) interior 
of the black hole.  In contrast, our method allows the singular region to
be approximately represented in the computational domain
\citep{BCCKV05,BCCKV06,CLMZ05,CLZ06}.  These new techniques have led to 
dramatically more effective numerical simulations of binary black hole systems, 
recently allowing us to determinate accurate waveforms for the final 
orbits and merger of equal mass systems \citep{BCCKV06}, which we 
generalize here for nonequal masses.
While some preliminary numerical calculations of recoil
from mergers of non-spinning black holes have been reported
\citep{Cam05,HSL06}, the initial separations have been too small
and the resolutions too coarse for reliable and precise numbers.

Here we report the first precise fully numerical estimates of the kick
received from the merger of two nonspinning black holes. We choose a 1.5:1
mass ratio because it is close to the analytically estimated optimal mass
ratio for maximum kick \citep{Fit83}, but is also close enough to equal mass that
resolution issues are not serious impediments to the numerical evolution.
In \S~2 we describe our numerical method and results, including convergence
tests.  We discuss the astrophysical implications of these results in \S~3.

\section{Numerical Simulations}

Reliable simulations of binary black hole mergers require the 
specification of constraint-satisfying initial field data followed by stable
evolution of the Einstein equations.  The evolution variables hold
information about gravitational fields in the form of tensor fields
representing the curvature of a vacuum spacetime.  Einstein's equations,
together with a crucial specification of gauge (i.e., coordinate) conditions,
then govern the evolution of these fields in time.

We utilize the ``puncture'' approach \citep{Brandt97b} to 
specify constraint-satisfying initial field configurations for black holes 
on an approximately circular inspiral trajectory within a few orbits 
of merger.  Specifically, we consider three initial separations 
of coordinate distances $d_{\rm init}=4.1M_0$, 
$d_{\rm init}=6.2M_0$, and $d_{\rm init}=7.0M_0$, where $M_0$ is the total initial gravitational mass 
of the system, and we use units where Newton's gravitational constant $G$ 
and the speed of light $c$ are set to unity
so that all quantities can be represented in terms of their mass-scaling. 
Further, we try several slight variations in the initial angular momentum for the largest 
separation.
These data begin our simulations as the system approaches the 
ISCO, allowing about 1-3 
orbits before the formation of a common horizon representing the final black 
hole.

This data is evolved by our finite-differencing code,
{\tt Hahndol} \citep{Imbiriba:2004tp}, 
on an adaptive mesh refinement structure
implemented via {\tt PARAMESH} \citep{MacNeice00}.  The punctures are allowed
to move freely through the grid and are evolved according to a variant
of the BSSN formulation of Einstein's equations and gauge conditions as
described in \citet{BCCKV06,VBKC06}.

We interpret our numerical results by studying the gravitational radiation
generated by the merger which carries gauge-invariant information away from 
the strong-field region.  In our simulations, radiation is represented by 
a component of the space-time Weyl curvature tensor,  $\psi_4$.
In terms of $\psi_4$, the time rate of change of radiated momentum 
can be expressed as follows \citep{NT80}:
\begin{eqnarray}\label{eq:Pdot}
\frac{dP_i}{dt}=\lim_{r \to \infty} \left\{ \frac{r^2}{4\pi}\int 
d\Omega \frac{x_i}{r} \left| \int_{-\infty}^t dt \psi_4  \right|^2 \right\}
\end{eqnarray}
We compute $\psi_4$ numerically, and extract it from the simulation data 
on a sphere of radius $r=50M_0$, 
which we have found is sufficiently large enough to give precise results.

\begin{figure}[t]
 \includegraphics*[scale=0.6, angle=-90]{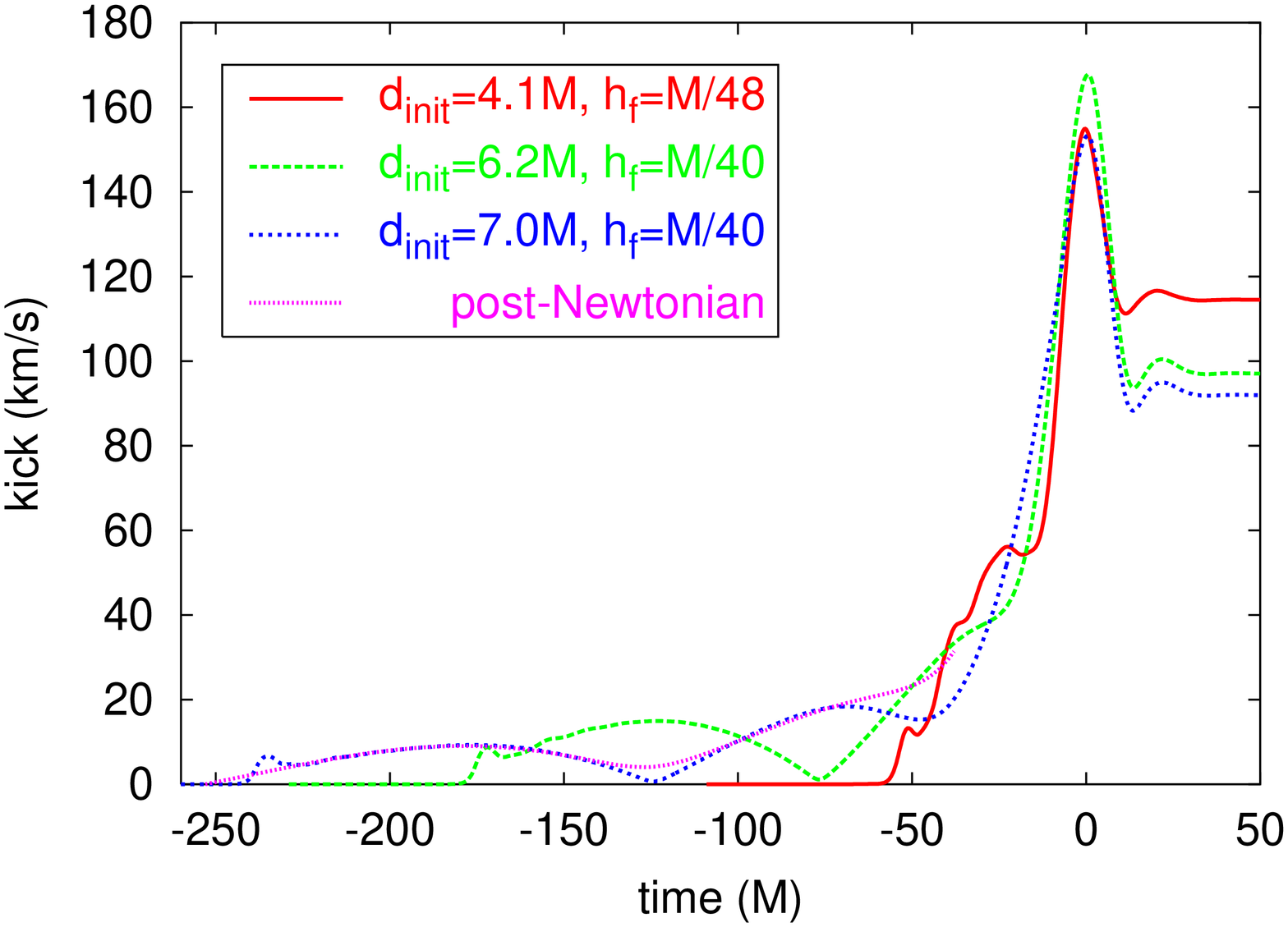}
 \caption{
The magnitude of the radiated momentum, as a function of time,
 from three different simulations.   
For initial coordinate separations of
$d_{\rm init}=4.1M_0, 6.2M_0, {\rm and}\ 7.0M_0$, the final
values of the kicks are respectively 113 km/s, 97 km/s, and 92 km/s.
 Also shown is the 2nd order
 post-Newtonian radiated momentum, which was computed from a
 low frequency cutoff commensurate with that of the $d_{\rm
 init}=7.0M_0$ simulation (see text for details). The excellent
 agreement of the post-Newtonian kick with that of the $d_{\rm
 init}=7.0M_0$ simulation over most of the first orbit,
 together with the agreement to within $6\%$ of the final kick from the $d_{\rm
 init}=7.0M_0$
 simulation with that of the $d_{\rm init}=6.2M_0$ simulation,
 lends support to the accuracy of these results.
 \label{fig:kickplot}}
\end{figure}

Our main numerical results are given in Figure~\ref{fig:kickplot},
which shows the net speed of the center of mass as a function of
time for several runs.  To compare runs at different initial coordinate 
separations we have shifted the time axis
so that the kick speed peaks at $t=0$.   Note that these curves are
representative of a much larger body of simulation data.  We have varied, for 
example, the finest grid spacing from $h_f=M_0/32$ to $h_f=M_0/48$ for the
$d_{\rm init}=4.1M_0$ runs, and from $h_f=M_0/32$ to $h_f=M_0/40$ for the 
$d_{\rm init}=7.0M_0$ runs, and in either set of runs found the kick results 
of different resolutions agree to within $2\%$, thus confirming our
numerical precision.  (And we have found that numerical error in these runs 
converges away at an acceptable rate; see \citep{BCCKV05,BCCKV06} for previous 
convergence 
tests of our code.)  We have also varied the extraction radius from $r=30M_0$ 
to $r=60M_0$ and, disallowing cases where the extraction radius significantly
intersects refinement boundaries, we have found agreement across extraction
radii to better than $2\%$.  We have also varied the initial, orbital angular 
momentum in the $d_{\rm init}=7.0M_0$ runs by as much as $4\%$, resulting
in a $6\%$ variation in the final kick values.  Our full range of final
kick values is 86-116~km~s$^{-1}$.   Excluding only the
$d_{\rm init}=4.1M_0$ case on the grounds that it starts with too little
initial separation, our
astrophysically relevant range of kick values is 86-97~km~s$^{-1}$.

The selection of our initial momenta and centers of mass was informed by
previous work on quasicircular initial data based on minimization of
an effective potential \citep{Pfeif05,PTC00,BCLT02,Cook94}, 
and further refined by trial and error.  The most astrophysically
relevant simulations were presumed to be those with the smoothest,
monotonic radiation frequencies and time derivatives thereof.  By these 
criteria our most reliable run gave a final kick of 92~km~s$^{-1}$. 

Also shown in Figure~\ref{fig:kickplot} is a post-Newtonian result
for the kick, as integrated from a low-frequency cutoff.  Because
our simulation starts from a finite orbit, the resulting radiation
has an effective low-frequency cutoff, which we have used in the 
post-Newtonian calculation for consistency.  We integrate, 
specifically, a 2PN formula for the orbital frequency \citep{BDIWW95}
and use the resulting frequency as a function of time in a
2PN formula for the radiated momentum \citep{BQW05}. 
For most of the first orbit, the agreement of the $d_{\rm init}=7.0M_0$ 
simulation represented in  Figure~\ref{fig:kickplot} with the 
post-Newtonian calculation is better than $1\%$.

Recently, two groups have made refined analytic estimates of the kick
from the merger of nonspinning black holes, both with precise answers but
differing in magnitude by a factor of three.   \citet{BQW05} predict a
speed of 155$\pm$25~km~s$^{-1}$ for a 1.5:1 mass ratio, whereas
\citet{DG06} predict 48~km~s$^{-1}$. Our result is between these estimates
but inconsistent with either one.  
We note, however, that Blanchet, Qusailah, \& Will (2005)
integrate only until the horizons overlap, which is close
to the peak of our kick.  At that point, our most reliable
run 
(i.e., with the widest initial separation and least apparent eccentricity)
gives a kick of $\approx 150$~km~s$^{-1}$, which is
encouragingly close to the analytic estimate.  This points
out the importance of the $\sim 40$\% post-peak reduction
in the kick.  This reduction can also be seen in Figure~5
of Damour \& Gopakumar (2006).  We attribute it to the
evolution by more than $\pi$ radians of the phase of the
emitted momentum during the merger, which thus partially
opposes the vector kick that exists at the time of horizon
overlap.  It is also noteworthy that, integrating only up to 
ISCO, Blanchet, Qusailah, \& Will (2005) 
obtain a kick of $\approx 14$~km~s$^{-1}$ for a mass ratio of 1.5:1, and in our 
most reliable run we also obtain
a kick of $\approx 14$~km~s$^{-1}$ at ISCO.

An initial numerical estimate for 1.04:1 
and 1.18:1 mass ratios was made by \citet{HSL06} (albeit at much lower
resolution and starting much closer than our simulations), and they find
kick speeds of 9~km~s$^{-1}$ and 33~km~s$^{-1}$, respectively.  
One can compare results at different mass with reference to the 
\citet{Fit83} fitting formula ${\rm kick}\propto
q^2(q-1)/(1+q)^5\,=\,(\mu/m)^2{\rm d}m/m$, where $q=m_1/m_2>1$ is the 
mass ratio, $\mu/m=m_1 m_2/(m_1+m_2)^2$ is the symmetric mass ratio
and ${\rm d}m/m=(m_1-m_2)/(m_1+m_2)$ is the fractional
mass difference.  
Although originally derived in the context of a leading order
post-Newtonian approximation, this formula is suggested to be a good
approximation by perturbative
Schwarzschild calculations \citep{FD84}, and it also closely agrees with 
more recent post-Newtonian results \citep{BQW05,DG06}.
By Fitchett's formula, 
the \citet{HSL06} results would
imply $\sim 80$~km~s$^{-1}$ for a mass ratio of 1.5:1, closer to our
final value than the analytic estimates.

\section{Discussion and Conclusions}

Our range of $101\pm 15$~km~s$^{-1}$ (and best estimate of
$92\pm 6$~km~s$^{-1}$) for the kick received
in a 1.5:1 merger of nonspinning black holes has important
implications for the assembly of supermassive black holes in
the early universe (see also \citealt{Mer04}).  This is
because as dark matter halos merge in the process of
hierarchical structure assembly, their central black holes
are also presumed to merge if the halo mass ratio is not too
extreme (otherwise dynamical friction on the halos is likely to be inefficient; 
see \citealt{Taf03}).  If black hole masses are linked to
the mass of their host halos, it is therefore expected that in the
early universe, black hole mergers are likely to involve
comparable-mass objects.  If the resulting kick exceeds the
escape speed of the merged dark matter halo, the halo
is left without a black hole.  This could, therefore, have
a significant impact on the number of black hole mergers
in the early universe.  In addition, if mergers between
comparable-mass halos are common at redshifts $z\gtrsim 10$,
these ejections might reduce substantially the fraction of halos
that host black holes.

A full appraisal of the consequences of recoil will require detailed 
numerical studies of the effects of black hole spin and mass ratio, which
will be our subject in future papers.  We can make initial
guesses by focusing on nonspinning black holes and by adopting as
before the \citet{Fit83} formula for the dependence of kick speed on
mass ratio. Setting $v_{\rm kick}(q=1.5)=92$~km~s$^{-1}$ fixes
the curve.


To estimate the escape speed from a dark matter halo, we follow
the treatment of \citet{Mer04}.  The escape speed
from a halo of virial mass $M_{\rm vir}$ is
$V_{\rm esc}^2=2c g(c)M_{\rm vir}/R_{\rm vir}$,
where as before we set $G=1$. 
Here $c$ is the concentration parameter for the halo,
$g(c)=\left[\ln(1+c)-c/(1+c)\right]^{-1}$, and we
define $R_{\rm vir}$ as
the radius inside which the average density is 
$\Delta_c=18\pi^2+82x-39x^2$ for a flat universe, where
$x\equiv \Omega_M(z)-1$ (see \citealt{BN98}, equation 6).
The critical density for a flat universe is $\rho_{\rm crit}
=3H_0^2/(8\pi)[\Omega_M(1+z)^3+\Omega_\Lambda]
\approx 10^{-29}$~g~cm$^{-3}(1+z)^3$ (e.g., \citealt{Pea99},
section 3.2).  We use $\Omega_M=0.27$
and $\Omega_\Lambda=0.73$.
>From equation (18) of \citet{Bul01}, the
concentration parameter scales as $c=9\left[M_{\rm vir}/(2\times
10^{13}\,
M_\odot)\right]^{-0.13}(1+z)^{-1}$.

\begin{figure}[t]
 \includegraphics*[scale=0.6, angle=-90]{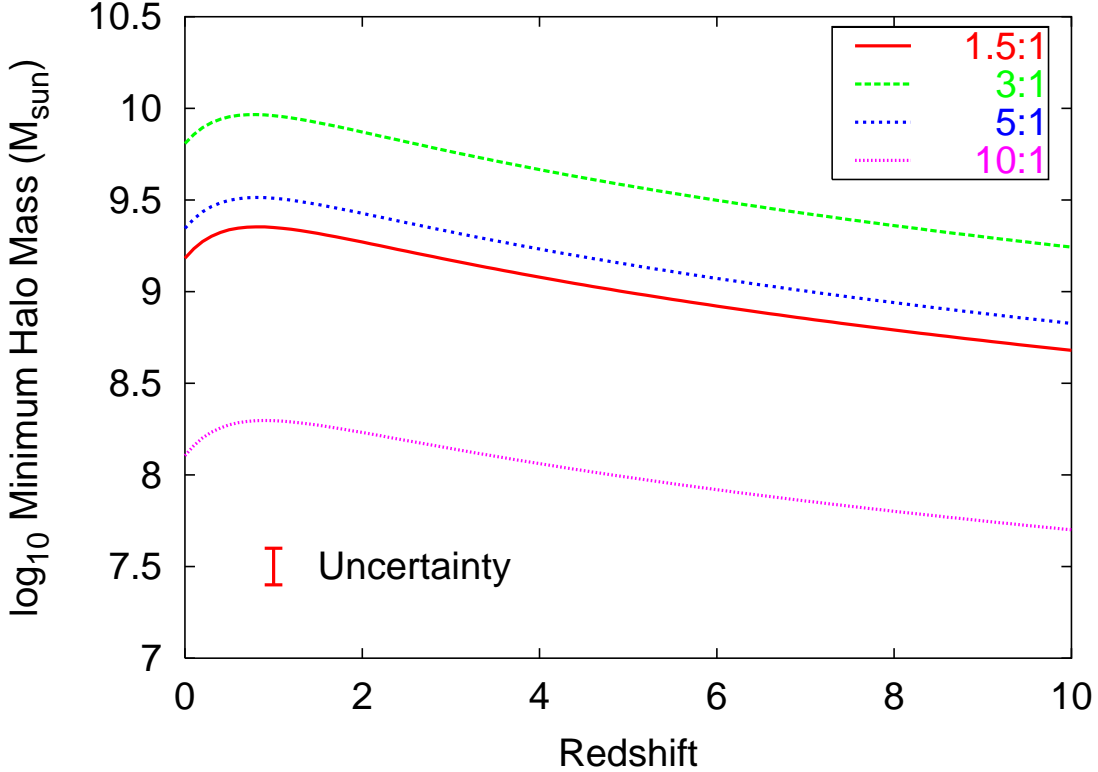}
 \caption{Minimum mass of a dark matter halo at a given redshift
 required to retain the product of the merger of two nonspinning
 black holes with a mass ratio indicated on the curve.  Details 
 of the computation are in the text.  Note that we use the
 \citet{Fit83} analytical estimate of the mass ratio dependence, 
 assuming our numerical result of 92~km~s$^{-1}$ for the mass ratio of 1.5:1.
 Then a mass ratio of 3:1 gives 140~km~s$^{-1}$, a mass ratio of 5:1 gives 
 103~km~s$^{-1}$, and a mass ratio of 10:1 gives 45~km~s$^{-1}$.
 This figure indicates that early halos might lose merger remnants
 because of the kick, but in the current universe only globular
 clusters or the smallest dwarf galaxies could have black holes ejected
 because of gravitational radiation recoil.
 \label{fig:escape}}
\end{figure}

In Figure~\ref{fig:escape} we plot the minimum halo mass as a function of
redshift such that $V_{\rm esc}>v_{\rm kick}$ for the listed
mass ratios of 1.5:1, 3:1, 5:1, and 10:1, as projected from our results
by the Fitchett scaling.  Note that the low
power of $M_{\rm vir}$ in the escape speed means that the
minimum halo mass depends sensitively on the kick speed.
For example, if the kick speed at a 1.5:1 mass ratio were
200~km~s$^{-1}$ instead of 92~km~s$^{-1}$, the threshold
halo mass at $z=10$ for this mass ratio would jump from 
$4\times 10^8\,M_\odot$ to $3\times 10^9\,M_\odot$.  This would in 
turn reduce the number density of halos massive enough to retain
their black holes by
a factor $\sim 20$ (see \citealt{MW02} for a pedagogical
discussion of how to estimate halo number densities).  This
points out the importance of estimating kick speeds accurately.

As discussed by \citet{Mer04}, kicks are also important
in the current-day universe for low-mass concentrations of
stars.  Their Figure~2 is a useful summary of escape speeds
from the centers of galaxies and globular clusters.  From their
figure, we see that comparable-mass mergers with kick speeds
$\sim 100$~km~s$^{-1}$ will cause ejection from globulars and
some dwarf galaxies, but that more massive galaxies will retain
the remnant.  Using the \citet{Fit83} scaling with mass ratio,
we find that black holes that are $\gtrsim 10$ times more
massive than their companions are not ejected from globular
clusters with escape speeds of a few tens of km~s$^{-1}$.
Coincidentally, this is approximately the same mass ratio
required to protect a massive black hole from cluster ejection
from three-body interactions \citep{GMH04,GMH06}.
Therefore, if an intermediate-mass black hole of $>100-200\,M_\odot$
is formed in a cluster, it can stay and
potentially grow through future interactions.
Stellar-mass binary black holes ($M<50\,M_\odot$) will be ejected from
such clusters prior to merger by Newtonian three-body interactions
\citep{KHM93,SH93}, hence recoil is not so important for low-mass
black holes in this context.  However, mergers of black hole binaries
in low-density galactic disks could produce a population of
high-speed coalesced black holes.

In conclusion, we have presented a reliable and precise,
fully numerical estimate of the gravitational recoil produced
by the merger of two unequal mass nonspinning black holes.
Our best estimate of 92$\pm$6~km~s$^{-1}$ for a 1.5:1 mass ratio is 
intermediate between the recent analytic estimates of \citet{BQW05}, who
suggest 155$\pm$25~km~s$^{-1}$ for this mass ratio, and
\citet{DG06}, whose formulae would imply 48~km~s$^{-1}$.
Our results are thus an important step in accurately evaluating
the astrophysical consequences of gravitational radiation recoil
in dense stellar clusters and the early universe.

\acknowledgments
We gratefully acknowledge supercomputing support from the Commodity
Cluster Computing Project (NASA-GSFC), 
the NASA Center for Computational Sciences (NASA-GSFC)
and Project Columbia (NASA Advanced Supercomputing Division, NASA-Ames).  
This work was supported in part by NASA grants ATP02-0043-0056 
and O5-BEFS-05-0044. JvM, MK, and MCM were also
supported in part by the Research Associateship Programs Office of the
National Research Council and Oak Ridge Associated Universities. 
Comments by Scott Hughes and Cliff Will on a earlier draft
helped clarify this manuscript.

\newpage

\newpage

\end{document}